\documentclass[12pt,leqno,a4paper]{article}
\usepackage{amsmath,amsthm,enumerate}
\usepackage[latin1]{inputenc}
\usepackage[T1]{fontenc}
\usepackage[english]{babel}

\newtheorem{theorem}{Theorem}[section]
\newtheorem{proposition}[theorem]{Proposition}
\newtheorem{lemma}[theorem]{Lemma}

\newtheorem{remark}[theorem]{Remark}


\newcommand*\proofnamestyle{\itshape}

\begin{document}

    \title{Characterization of symmetric monotone metrics on the state space of quantum systems}
      \author{Frank Hansen}
      \date{January 26th, 2006\\
      {\tiny Revision March 28th, 2006}}
      \maketitle

      \begin{abstract}

      The quantum Fisher information is a Riemannian metric, defined on the state space of a quantum system,
      which is symmetric and decreasing under stochastic mappings. Contrary to
      the classical case such a metric is not unique. 
      We complete the characterization, initiated by Morozova, Chentsov and Petz, of these metrics 
      by providing a closed and tractable
      formula for the set of Morozova-Chentsov functions. In addition, we provide a continuously
      increasing bridge between the smallest and largest symmetric monotone metrics.
      \end{abstract}

    \section{Introduction}

    In the geometric approach to classical statistics the canonical Riemannian metric is given by the Fisher
    information, and it measures the statistical distinguishability of probability distributions.
    The Fisher metric is the unique Riemannian metric contracting under
    Markov morphisms \cite{kn:censov:1982}.

    In quantum mechanics the probability simplex is replaced by the state space of density matrices 
    (positive semi-definite trace one matrices), and Markov
    morphisms are replaced by stochastic mappings. A linear map $ T\colon M_n(\mathbf C)\to M_m(\mathbf C) $
    is said to be stochastic if it is completely positive and trace preserving. Since stochastic mappings 
    (like their Markovian counter parts)
    represent coarse graining or randomization, one would expect statistical distinguishability of states to
    decrease under stochastic mappings.

    These considerations lead Chentsov and Morozova to define a monotone metric as a map (or rather a family of maps)
    $ \rho\to K_\rho $ from the set $ \mathcal M_n $ of positive definite $ n\times n $ density matrices
    to sesquilinear\footnote{We use the complexification proposed by
    Petz \cite{kn:petz:1996:2}.}
    forms $ K_\rho(A,B) $ defined on $ M_n(\mathbf C) $ satisfying:
    \begin{enumerate}[(i)]

    \item $ K_\rho(A,A)\ge 0, $ and equality holds if and only if $ A=0. $

    \item $ K_\rho(A,B)=K_\rho(B^*,A^*) $ for all $ \rho\in\mathcal M_n $ and all $ A,B\in M_n(\mathbf C). $

    \item $ \rho\to K_\rho(A,A) $ is continuous on $ \mathcal M_n $ for every $ A\in M_n(\mathbf C). $

    \item $ K_{T(\rho)}(T(A),T(A))\le K_\rho(A,A) $ for every $ \rho\in\mathcal M_n, $ every
    $ A\in M_n(\mathbf C) $ and every stochastic mapping $ T:M_n(\mathbf C)\to M_m(\mathbf C). $

    \end{enumerate}

    It is understood that these requirements should hold for all $ n $ and $ m. $ The condition (ii)
    is sometimes omitted, but we shall only consider symmetric metrics. Since condition (iv) implies unitary
    covariance we may in all calculations assume that $ \rho $ is a diagonal matrix. Chentsov and Morozova
    proved that there to each monotone metric $ K $ is a positive function $ c(\lambda,\mu) $
    defined in the first quadrant and a positive constant $ C $ such that
    \begin{gather}\label{metric in terms of a Morozova-Chentsov function}
    K_\rho(A,A)=C\sum_{i=1}^n \lambda_i^{-1}|A_{ii}|^2 + \sum_{i\ne j}|A_{ij}|^2 c(\lambda_i,\lambda_j)
    \end{gather}
    for each diagonal matrix $ \rho\in\mathcal M_n $ with diagonal $ (\lambda_1,\dots,\lambda_n) $
    and every $ A $ in $ M_n(\mathbf C). $ The metric is therefore fully described by the
    so called Morozova-Chentsov function $ c $ which is symmetric in its two variables and satisfy
    \[
    c(\lambda,\lambda)=C\lambda^{-1}\quad\text{and}\quad c(t\lambda,t\mu)=t^{-1}c(\lambda,\mu)
    \]
    for all $ t,\lambda,\mu>0. $
    At the time it was completely unsettled which type of functions $ c $ would, through the formula
    (\ref{metric in terms of a Morozova-Chentsov function}), give rise to a monotone metric.
    It was not even clear that there existed
    a single Morozova-Chentsov function (and thus a monotone metric) although Morozova and Chentsov had
    a few candidates.

    Petz connected the theory of monotone metrics with the theory of connections and means by Kubo and
    Ando \cite{kn:kubo:1980} and was able to prove that the set of Morozova-Chentsov functions are given
    on the form
    \begin{gather}\label{Morozova-Chentsov function}
    c(\lambda,\mu)=\frac{1}{\mu f(\lambda\mu^{-1})}\qquad \lambda,\mu >0,
    \end{gather}
    where $ f $ is a positive operator monotone function defined on the positive half-axis satisfying the
    functional equation
    \begin{gather}\label{functional equation}
    f(t)=tf(t^{-1})\qquad t>0.
    \end{gather}
    Petz was by this result able to give some examples of Morozova-Chentsov functions, and they included the
    candidates put forward by Morozova and Chentsov. The existence of monotone metrics was then established.

    It is however a problem that the class of operator monotone functions satisfying (\ref{functional equation})
    is largely unknown. The aim of the present paper is to provide a closed and tractable formula for the
    set of Morozova-Chentsov functions and in this way complete the characterization of (symmetric)
    monotone metrics given by Morozova, Chentsov and Petz. The main result is the formula given in
    Theorem~\ref{theorem: set of Morozova-Chentsov functions}. In addition, we provide
    a continuously increasing bridge (\ref{bridge of Morozova-Chnetsov functions})
    between the smallest and largest (symmetric) monotone metrics.

    \section{Statement of the main results}

      Let $ f $ be a positive operator monotone function defined
      on the positive half-axis. It has a canonical representation \cite{kn:kubo:1980} of the form
      \begin{gather}\label{formula for operator monotone function}
      f(t)=\int_0^\infty \frac{t(1+s)}{t+s}\,d\mu(s)\qquad t>0,
      \end{gather}
      where $ \mu $ is a positive (non-vanishing) finite measure on the extended half-line $ [0,\infty]. $
      The function $ f^\# $ defined by setting
      \[
      f^\#(t)=t f(t^{-1})\qquad t\in\mathbf R_+
      \]
      is operator monotone. This follows easily from the above integral representation, but may also be
      inferred by much simpler algebraic arguments \cite[2.1.~Theorem (v) and 2.5~Theorem]{kn:hansen:1982}
      without the use of Löwner's deep theory.
      Since $ f^{\#\#}=f $ the operation $ f\to f^\# $ is an involution on the set
      of positive operator monotone functions defined on the positive half-axis. The harmonic mean is
      separately (operator) increasing, hence also the function
      \begin{gather}\label{definition: tilde f}
      \tilde f(t)=H(f(t),f^\#(t))=\frac{2f(t)f^\#(t)}{f(t)+f^\#(t)}
      \end{gather}
      is operator monotone. It is an easy calculation to show that $ (\tilde f)^\# = \tilde f. $
      The formula (\ref{definition: tilde f}) therefore associates a positive operator monotone
      function $ \tilde f $ satisfying the functional equation (\ref{functional equation}) to any positive
      operator monotone function $ f, $ and $ \tilde f= f $ if already $ f^\#=f. $ This procedure is
      implicitly applied in \cite[Formula (12)]{kn:petz:1996:2} where Petz calculates a Morozova-Chentsov function
      from an operator monotone function not necessarily satisfying the functional
      equation (\ref{functional equation}).

      There are several problems with this method, although it may be useful to calculate explicit examples.
      Firstly, the mapping $ f\to\tilde f $ is not injective. There are
      in general infinitely many operator monotone functions which are mapped to the same function.
      Secondly, it seems difficult to specify the induced equivalence relation on the set of positive finite
      measures on the extended half-line $ [0,\infty]. $

    \begin{theorem}\label{theorem: canonical representation of f}
    Let $ f\colon\mathbf R_+\to\mathbf R_+ $ be a function satisfying
    \begin{enumerate}[(i)]
    \item $ f $ is operator monotone,
    \item $ f(t)=tf(t^{-1}) $ for all $ t>0. $
    \end{enumerate}
    Then $ f $ admits a canonical representation
    \begin{gather}\label{canonical representation of f}
    f(t)=e^\beta\frac{1+t}{\sqrt{2}}\exp\int_0^1\frac{\lambda^2-1}{\lambda^2+1}\cdot
    \frac{1+t^2}{(\lambda+t)(1+\lambda t)}h(\lambda)\,d\lambda
    \end{gather}
    where $ h:[0,1]\to[0,1] $ is a measurable function and $ \beta $ is a real constant.
    Both $ \beta $ and the equivalence class containing $ h $
    are uniquely determined by $ f. $ Any function
    on the given form maps the positive half-axis into itself and satisfy $ (i) $ and $ (ii). $
    \end{theorem}

    \begin{proof}
    The result follows by applying Lemma \ref{fundamental lemma}
    and Theorem \ref{function in E satisfying the functional equation}.
    \end{proof}

    Note that $ \exp\beta=f(i)\exp(-i\pi/4). $ We may adjust the constant $ \beta $ such that $ f(1)=1. $
    This corresponds to setting the constant $ C=1 $
    in formula (\ref{metric in terms of a Morozova-Chentsov function}) and gives rise to a
    so called Fisher adjusted metric.
    We are now able to calculate the set of Morozova-Chentsov functions.

    \begin{theorem}\label{theorem: set of Morozova-Chentsov functions}
    A Morozova-Chentsov function $ c $ admits a canonical representation
    \begin{gather}\label{canonical representation of c}
    c(x,y)=\frac{C_0}{x+y}
    \exp\int_0^1\frac{1-\lambda^2}{\lambda^2+1}\cdot
    \frac{x^2+y^2}{(x+\lambda y)(\lambda x +y)}h(\lambda)\,d\lambda
    \end{gather}
    where $ h:[0,1]\to[0,1] $ is a measurable function and $ C_0 $ is a positive constant.
    Both $ C_0 $ and the equivalence class containing $ h $
    are uniquely determined by $ c. $ Any function $ c $
    on the given form is a Morozova-Chentsov function.
    \end{theorem}

    Note that $ c $ is increasing in $ h $ and that the constant $ C_0 $ may be adjusted such that $ c(x,x)=x^{-1}. $
    For Morozova-Chentsov functions $ c_h $ with a fixed constant $ C_0 $ and $ h $ as 
    in formula (\ref{canonical representation of c}) we have $ c_{sh+(1-s)g}=c_h^s c_g^{1-s}, $ 
    $ 0\le s\le 1. $ 

    \begin{proposition}
    Let the exponent $ \gamma\in[0,1]. $ The functions
    \[
    f_\gamma(t)=\frac{1}{2} (1+t)\left(\frac{4t}{(t+1)^2}\right)^{\gamma}
    =t^\gamma\left(\frac{1+t}{2}\right)^{1-2\gamma}\qquad t>0
    \]
    are operator monotone, normalized in the sense that $ f(1)=1 $ and satisfy the functional equation
    $ f(t)=t f(t^{-1}) $ for $ t>0. $
    \end{proposition}

    \begin{proof} We first calculate the integral
    \[
    \int_0^1\frac{\lambda^2-1}{\lambda^2+1}\cdot
    \frac{1+t^2}{(\lambda+t)(1+\lambda t)}\,d\lambda
    =\log\frac{2t}{(1+t)^2}
    \]
    and then by setting $ h(\lambda)=\gamma $ in (\ref{canonical representation of f}) obtain
    the operator monotone function
    \[
    f(t)=e^\beta \frac{1+t}{\sqrt{2}}\left(\frac{2t}{(1+t)^2}\right)^\gamma
    \]
    satisfying the functional equation (\ref{functional equation}).  The result now follows by
    setting $ \beta=(\gamma-1/2)\log 2. $
    \end{proof}

    \begin{remark}\rm
    Once we found the functions $ f_\gamma $ above, we may directly verify that they
    are operator monotone and satisfy the functional equation (\ref{functional equation}) without making
    use of Theorem~\ref{theorem: canonical representation of f}. Indeed, it is trivial that they satisfy
    equation (\ref{functional equation}).
    To see that they are operator monotone we take a complex number $ z=r\exp(i\theta) $
    in the upper half plane, that is $ 0<\theta<\pi. $ Since $ 1+z $ is translated one unit to the right as
    compared with $ z $ it is still located in the upper half plane,
    but the angle with the real axis has decreased.
    It can therefore be written on the form $ 1+z=r_1\exp(i\theta_1) $ where $ 0<\theta_1<\theta $
    and we notice that $ 0<\theta-\theta_1<\theta<\pi. $
    The analytic continuation of $ f_\gamma $ to $ z $ may thus be written on the form
    \[
    f_\gamma(z)=\frac{4^\gamma r^\gamma r_1^{1-2\gamma}}{2}\exp\bigl(i(\gamma\theta + (1-2\gamma)\theta_1)\bigr).
    \]
    But since $ \gamma\theta + (1-2\gamma)\theta_1=\gamma(\theta-\theta_1)+(1-\gamma)\theta_1\in (0,\pi), $
    we derive that the imaginary part of $ f(z) $ is positive. But this proves that $ f_\gamma $ is operator
    monotone by Löwner's theorem \cite{kn:loewner:1934, kn:donoghue:1974}.
    
    A third proof is obtained by considering the set $ E $ of exponents $ \gamma\in[0,1] $ such that $ f_\gamma $
    is operator monotone. Since $ f_{(\gamma+\delta)/2}=(f_\gamma f_\delta)^{1/2} $ and the geometric mean is
    operator increasing, we derive that $ E $ is mid-point convex, and since $ E $ is closed
    and contains $ 0 $ and $ 1, $ we obtain $ E=[0,1]. $ 
    \end{remark}

    The set of operator monotone functions $ f $ defined on
    the positive half-axis such that $ f(1)=1 $ and $ f(t)=tf(t^{-1}) $ for all $ t>0 $
    has a minimal and a maximal element \cite{kn:petz:1996, kn:kubo:1980}. These extremal functions are given by
    \[
    f_1(t)=\frac{2t}{1+t}\;\text{(min)}\quad\text{and}\quad f_0(t)=\frac{1+t}{2}\;\text{(max)}.
    \]
    Since $ 4t(t+1)^{-2}\le 1 $ for all $ t>0 $ we deduce that
    \begin{gather}
    0\le\gamma\le\delta\le 1\quad\Rightarrow\quad f_\gamma(t)\ge f_\delta(t)\quad\forall t>0.
    \end{gather}
    The family $ (f_\gamma(t))_{\gamma\in [0,1]} $ therefore provides a continuously decreasing bridge between the
    above extremal functions.  Note also that $ f_{1/2}(t)=t^{1/2}. $
    The Morozova-Chentsov functions corresponding to the family $ (f_\gamma(t))_{\gamma\in [0,1]} $ are given by
    \begin{gather}\label{bridge of Morozova-Chnetsov functions}
    c_\gamma(x,y)=x^{-\gamma} y^{-\gamma}\left(\frac{x+y}{2}\right)^{2\gamma-1}\qquad\gamma\in[0,1],
    \end{gather}
    and they provide a continuously increasing bridge between the smallest and largest symmetric monotone metrics.
    Finally, since $ c_{s\gamma+(1-s)\delta}=c_\gamma^s c_\delta^{1-s} $ for $ \gamma,\delta,s\in[0,1] $ we realize
    that the mapping $ \gamma\to c_\gamma $ is $ \log $-affine.

    \section{The exponential order relation}

    We introduced in an earlier paper \cite{kn:hansen:1981} the exponential ordering $ \preceq $ between
    linear self-adjoint operators on a Hilbert space by setting $ A\preceq B $ if $ \exp A\le\exp B. $
    It is easily verified that $ \preceq $ is an order relation, and since the logarithm is operator monotone
    it follows that $ A\preceq B $ implies $ A\le B. $ This is expressed by saying that
    the order relation $ \preceq $ is stronger than $ \le. $

    We also introduced and studied the set $ \mathcal E $
    of real functions $ F:\mathbf R\to\mathbf R $ which are monotone with
    respect to the exponential ordering.

    \begin{proposition}\label{proposition: fundamental bijection}
    The mapping $ \Phi $ defined by setting
    \begin{gather}
    \Phi(F)(t)=\exp F(\log t)\qquad x\in\mathbf R
    \end{gather}
    is a bijection of $ \mathcal E $ onto the set $ \mathcal P $ of
    positive operator monotone functions defined on the positive half-axis.
    \end{proposition}

    \begin{proof} Let $ A $ and $ B $ be positive invertible operators. Then
    \[
    \begin{array}{rl}
    \Phi(F)(A)\le\Phi(F)(B) &\Leftrightarrow\: \exp F(\log A)\le\exp F(\log B)\\[1ex]
    &\Leftrightarrow\: F(\log A)\preceq F(\log B),
    \end{array}
    \]
    and the assertion follows since the logarithm maps the positive half-line onto the real line.
    \end{proof}

    The next result was proved in \cite[Theorem 2.3]{kn:hansen:1981}.

    \begin{theorem}
    A non-constant function $ F:\mathbf R\to\mathbf R $ belongs to the class
    $ \mathcal E $ if and only if it admits an analytic continuation into
    the strip $ \{z\in\mathbf C\mid 0<\Im z<\pi\} $ which leaves the strip
    invariant.
    \end{theorem}

    Based on this result and by applying the theory of analytic functions we
    obtained \cite[Theorem 2.4]{kn:hansen:1981} the following representation theorem.

    \begin{theorem}
    A function $ F\colon\mathbf R\to\mathbf R $ is in the class $ \mathcal E $
    if and only if it admits a canonical representation
    \begin{gather}\label{canonical representation of F}
    F(x)=\beta+\int_{-\infty}^0\left(\frac{1}{\lambda-\exp x} -
    \frac{\lambda}{\lambda^2+1}\right) h(\lambda)\,d\lambda\qquad x\in\mathbf R,
    \end{gather}
    where $ h:(-\infty,0]\to [0,1] $ is a measurable function and $ \beta $ is a real constant.
    The constant $ \beta $ and the equivalence class containing $ h $ are uniquely determined by $ F. $
    \end{theorem}

    \section{The functional equation}

    The next result is the key observation in the present article.

    \begin{lemma}\label{fundamental lemma}
    Let $ F $ be a function in $ \mathcal E. $ The function
    $ f=\Phi(F)\in\mathcal P $ satisfies the functional equation (\ref{functional equation})
    if and only if $ F(x)=x+F(-x) $ for every $ x\in\mathbf R. $
    \end{lemma}

    \begin{proof}
    By taking the logarithm in the equation
    \[
    \exp F(\log t)=\Phi(F)(t)=f(t)=tf(t^{-1})=t\exp F(\log t^{-1}),
    \]
    we realize that the functional equation (\ref{functional equation}) for $ f $ is equivalent to
    \[
    F(\log t)=\log t + F(-\log t),
    \]
    or to $ F(x)=x+F(-x) $ by setting $ x=\log t. $
    \end{proof}

    We need the following lemma as a preparation to the main theorem in this section.

    \begin{lemma}\label{definite integral}
    \[
    \int_{-1}^0 \frac{2\sin\theta}{\lambda^2-2\lambda\cos\theta +1}\,d\lambda=\theta\qquad 0<\theta<\pi.
    \]
    \end{lemma}

    \begin{proof}
    Since the integrand can be written as $ 2\Im (\lambda-e^{i\theta})^{-1} $
    the integral is calculated to be
    \[
    \begin{array}{rl}
    2\Im\left[\log(\lambda-e^{i\theta})\right]_{-1}^0&\displaystyle
    =2\Im\log\frac{e^{i\theta}}{1+e^{i\theta}}=2\Im\log\frac{e^{i\theta/2}}{2\cos(\theta/2)}\\[2.5ex]
    &=2\Im\left(-\log 2\cos(\theta/2) + \log e^{i\theta/2}\right)\\[2ex]
    &=\theta,
    \end{array}
    \]
    where we used the complex logarithm.
    \end{proof}

    \begin{theorem}\label{theorem: functional equation for h}
    A function $ F\in\mathcal E $ with canonical representation as given
    by (\ref{canonical representation of F}) satisfies the functional
    equation
    \[
    F(x)=x+F(-x)\qquad\forall x\in\mathbf R,
    \]
    if and only if $ h(\lambda^{-1})=1-h(\lambda) $ for almost all $ \lambda\in [-1,0). $
    \end{theorem}

    \begin{proof}
    Suppose that a function $ F\in\mathcal E $ satisfies the given
    functional equation. Applying analytic continuation into the strip $ \{z\in\mathbf C\mid 0<\Im z<\pi\} $ and
    setting $ x=i\theta, $ we thus obtain
    \begin{gather}\label{functional equation for imaginary numbers}
    F(i\theta)=i\theta + F(-i\theta)\qquad 0<\theta<\pi.
    \end{gather}
    Inserting the integral expression (\ref{canonical representation of F}) we then get
    \[
    \begin{array}{rl}
    F(i\theta)-F(-i\theta)&=\displaystyle\int_{-\infty}^0
    \left(\frac{1}{\lambda-\exp(i\theta)}-\frac{1}{\lambda-\exp(-i\theta)}\right)h(\lambda)\\[3ex]
    &\displaystyle=\int_{-\infty}^0\frac{2i\sin\theta}{\lambda^2-2\lambda\cos\theta+1}h(\lambda)\,d\lambda\\[3ex]
    &=i\theta,
    \end{array}
    \]
    or equivalently
    \begin{gather}\label{integral to be split}
    \int_{-\infty}^0\frac{2\sin\theta}{\lambda^2-2\lambda\cos\theta+1}h(\lambda)\,d\lambda=\theta\qquad
    0<\theta<\pi.
    \end{gather}
    We split the range of integration at the point $ \lambda=-1 $ and make the variable change
    $ \lambda\to\lambda^{-1} $ in the first term and calculate
    \[
    \begin{array}{rl}
    \theta&=\displaystyle\int_{-\infty}^0
    \frac{2\sin\theta}{\lambda^2-2\lambda\cos\theta+1}h(\lambda)\,d\lambda\\[3ex]
    &=\displaystyle\int_0^{-1}
    \frac{2\sin\theta}{\lambda^{-2}-2\lambda^{-1}\cos\theta+1}h(\lambda^{-1})\frac{-1}{\lambda^2}\,d\lambda
    +\displaystyle\int_{-1}^0\frac{2\sin\theta}{\lambda^2-2\lambda\cos\theta+1}h(\lambda)\,d\lambda.
    \end{array}
    \]
    By applying Lemma \ref{definite integral} we therefore obtain
    \[
    \int_{-1}^0\frac{2\sin\theta}{\lambda^2-2\lambda\cos\theta+1}\bigl(h(\lambda)+h(\lambda^{-1})-1\bigr)\,d\lambda
    =0\qquad 0<\theta<\pi,
    \]
    which is simplified to
    \[
    \int_{-1}^0\frac{1}{\lambda^2-2\lambda\cos\theta+1}\bigl(h(\lambda)+h(\lambda^{-1})-1\bigr)\,d\lambda=
    0\qquad 0<\theta<\pi.
    \]
    We introduce the change of variable $ u=2\lambda(\lambda^2+1)^{-1} $ and note that $ u(-1)=-1 $ and $ u(0)=0. $
    Since the derivative
    \[
    u'(\lambda)=\frac{2(1-\lambda^2)}{(\lambda^2+1)^2}>0\qquad -1<\lambda\le 0,
    \]
    we may write $ \lambda=\lambda(u) $ as an increasing function of $ u. $ By introducing the function
    \[
    g(\lambda)=\frac{\lambda^2+1}{2(1-\lambda^2)}(h(\lambda)+h(\lambda^{-1})-1)\qquad -1\le\lambda< 0,
    \]
    we may write the above integral on the form
    \[
    \int_{-1}^0\frac{1}{1-2\lambda(\lambda^2+1)^{-1}\cos\theta}\, g(\lambda)u'(\lambda)\, d\lambda,
    \]
    hence
    \[
    \int_{-1}^0\frac{1}{1-u\cos\theta}\, g(\lambda(u))\, du=0\qquad 0<\theta<\pi.
    \]
    Setting $ t=\cos\theta $ we obtain that the function
    \[
    \varphi(t)=\sum_{n=0}^\infty t^n\int_{-1}^0 u^ng(\lambda(u))\,du=0\qquad -1<t<1,
    \]
    hence the derivatives
    \[
    \varphi^{(n)}(0)=n!\int_{-1}^0 u^n g(\lambda(u))\,du=0\qquad n=0,1,2,\dots.
    \]
    We conclude that the function $ u\to g(\lambda(u)) $ vanish for almost all $ u, $ and
    since $ \lambda\to u(\lambda) $ maps sets with positive Lebesgue measure to sets with
    positive Lebesgue measure, we derive that $ g(\lambda)=0 $ for almost all $ \lambda\in[-1,0]. $
    But this shows that $ h(\lambda^{-1})=1-h(\lambda) $ for almost all $ \lambda\in [-1,0). $

    If on the other hand this relationship is assumed, we may calculate backwards and obtain that $ F $ satisfies
    the functional equation (\ref{functional equation for imaginary numbers}). The assertion then follows
    by applying analytic continuation and setting $ \theta=-ix. $
    \end{proof}

    \begin{theorem}\label{function in E satisfying the functional equation}
    A function $ F\colon\mathbf R\to\mathbf R $ is in the class $ \mathcal E $ and satisfy the functional equation
    $ F(x)=x+F(-x) $
    for all $ x\in\mathbf R $ if and only if it admits a canonical representation
     \[
    F(x)=\beta+\log\frac{1+\exp x}{\sqrt{2}}+\int_0^1\frac{\lambda^2-1}{\lambda^2+1}\cdot
    \frac{1+\exp2x}{(\lambda+\exp x)(1+\lambda\exp x)}h(\lambda)\,d\lambda
    \]
    where $ h:[0,1]\to[0,1] $ is a measurable function and $ \beta\in\mathbf R. $
    The equivalence class containing $ h $ is uniquely determined by $ F, $ and $ \beta=\Re F(i\pi/2). $
    \end{theorem}

    \begin{proof}
    Take a function $ F\in\mathcal E $ with canonical representation as given
    by (\ref{canonical representation of F}) and satisfying the functional equation.
    Then $ h(\lambda^{-1})=1-h(\lambda) $ for almost all
    $ \lambda\in [-1,0) $ by Theorem \ref{function in E satisfying the functional equation}.
    By splitting the integral at the point $ \lambda=-1 $ in the integral representation
    (\ref{canonical representation of F}) and making the substitution $ \lambda\to\lambda^{-1} $
    in the first term, we obtain
    \[
    \begin{array}{rl}
    F(x)&=\displaystyle\beta-\int_0^{-1}\left(\frac{1}{\lambda^{-1}-\exp x} -
    \frac{\lambda^{-1}}{\lambda^{-2}+1}\right) h(\lambda^{-1})\,\frac{d\lambda}{\lambda^2}\\[3ex]
    &\hskip 9em\displaystyle +\int_{-1}^0\left(\frac{1}{\lambda-\exp x} -
    \frac{\lambda}{\lambda^2+1}\right) h(\lambda)\,d\lambda\\[3ex]
    &\displaystyle=\beta+\int_{-1}^0\left(\frac{\lambda^{-1}}{1-\lambda\exp x} -
    \frac{\lambda^{-1}}{1+\lambda^2}\right) (1-h(\lambda))\,d\lambda\\[3ex]
    &\hskip 9em\displaystyle +\int_{-1}^0\left(\frac{1}{\lambda-\exp x} -
    \frac{\lambda}{\lambda^2+1}\right) h(\lambda)\,d\lambda,
    \end{array}
    \]
    where we used Theorem \ref{theorem: functional equation for h}.
    Consequently
    \[
    \begin{array}{l}
    F(x)=\displaystyle\beta+\displaystyle\int_{-1}^0\left(\frac{\lambda^{-1}}{1-\lambda\exp x} -
    \frac{\lambda^{-1}}{\lambda^2+1}\right)\,d\lambda\\[3ex]
    \hskip 6em+\displaystyle\int_{-1}^0\left(\frac{1}{\lambda-\exp x} - \frac{\lambda^{-1}}{1-\lambda\exp x}
    -\frac{\lambda}{\lambda^2+1}+\frac{\lambda^{-1}}{1+\lambda^2}\right) h(\lambda)\,d\lambda,
     \end{array}
     \]
     and since
     \[
     \begin{array}{l}
     \displaystyle\int_{-1}^0\lambda^{-1}\left(\frac{1}{1-\lambda\exp x} - \frac{1}{1+\lambda^2}\right)\,d\lambda
     =\int_0^1\frac{\exp x-\lambda}{(1+\lambda\exp x)(1+\lambda^2)}\,d\lambda\\[3ex]
     =\displaystyle\Bigl[\log(1+\lambda\exp x)-\frac{1}{2}\log(1+\lambda^2)\Bigr]_{\lambda=0}^{\lambda=1}\\[2ex]
     =\displaystyle \log(1+\exp x)-\frac{1}{2}\log 2=\log\frac{1+\exp x}{\sqrt{2}}
     \end{array}
     \]
     we obtain
     \[
    F(x)=\beta+\log\frac{1+\exp x}{\sqrt{2}}+\int_{-1}^0\frac{1-\lambda^2}{1+\lambda^2}\cdot
    \frac{1+\exp2x}{(\lambda-\exp x)(1-\lambda\exp x)}h(\lambda)\,d\lambda.
    \]
    Defining $ h:[0,1]\to[0,1] $ by setting $ h(\lambda)=h(-\lambda) $ we may write
    \[
    F(x)=\beta+\log\frac{1+\exp x}{\sqrt{2}}+\int_0^1\frac{\lambda^2-1}{\lambda^2+1}\cdot
    \frac{1+\exp2x}{(\lambda+\exp x)(1+\lambda\exp x)}h(\lambda)\,d\lambda
    \]
    which is the desired expression. Calculating backwards we first extend
    $ h $ to the interval $ [-1,0] $ by setting $ h(-\lambda)=h(\lambda), $ and then to the interval $ ]-\infty,0] $ 
    by setting $ h(\lambda^{-1})=1-h(\lambda). $
    We arrive in this way at the integral expression (\ref{canonical representation of F}), and the sufficiency
    thus follows by Theorem \ref{theorem: functional equation for h}.
    \end{proof}

    {\footnotesize



      \vfill

      \noindent Frank Hansen: Department of Economics, University
       of Copenhagen, Studiestraede 6, DK-1455 Copenhagen K, Denmark.}

      \end{document}